\documentclass[useAMS,usenatbib]{mn2e}

\usepackage{epsfig}
\usepackage{graphicx}
\usepackage{amsmath} 
\usepackage{amssymb}
\usepackage{myaasmacros}

\newcommand{\beq}{\begin{equation}}
\newcommand{\eeq}{\end{equation}}
\def\LCDM{$\Lambda\mbox{CDM}$ }

\def\kms{{\rm km s}^{-1}}
\def\cm{{\rm\thinspace cm}}

\def\Msun{\hbox{$\thinspace \rm M_{\odot}$}}
\def\Msunh{\hbox{$\thinspace  \it h^{\rm -1} \rm \thinspace M_{\odot} $}}

\def\pch{{\rm\thinspace pc} \thinspace h^{\rm -1}}

\def\yr{{\rm\thinspace yr}}
\def\pdot{\dot{p}}

\def\Edotw{\dot{E}_{\rm w}}
\def\eps{\epsilon}
\def\epsw{\eps_{\rm f}}

\def\Mvir{M_{\rm vir}}
\def\Mdotacc{\dot{M}_{\rm acc}}
\def\Mdotinf{\dot{M}_{\rm inf}}
\def\Mdotoutf{\dot{M}_{\rm outf}}

\def\rvir{r_{\rm vir}}
\def\re{r_{\rm eff}}
\def\r10{r_{\rm 10}}

\def\Mbh{M_{\rm BH}}
\def\Mstel{M_{\rm stel}}

\def\sigmastar{\sigma}
\def\vw{v_{\rm w}}
\def\kms{\rm km~s^{-1}}
\def\mEdd{m_{\rm Edd}}
\def\ergs{\rm erg~s^{-1}}
\def\tH{t_{\rm H}}
\def\yr{\rm yr^{-1}}

\title[AGN feedback of early-type galaxies]
{The impact of mechanical AGN feedback on the formation of massive early-type galaxies}

\author[Choi et al.]{\parbox[t]{\textwidth}{
       Ena Choi$^{1,2}$\thanks{E-mail:enachoi@physics.rutgers.edu}, 
              Jeremiah P. Ostriker$^{2,3}$,
              Thorsten Naab$^{4}$,
              Ludwig Oser$^{3}$,
               Benjamin P. Moster$^{5}$}  
                \vspace*{6pt} \\  
        $^1$Department of Physics and Astronomy, Rutgers, The 
        State University of New Jersey, Piscataway, NJ 08854, USA\\
        $^2$Department of Astrophysical Sciences, Princeton
         University, Princeton, NJ 08544, USA \\
        $^3$Department of Astronomy, Columbia University, New York, 
        NY 10027, USA \\
        $^4$Max-Planck-Institut f\"ur Astrophysik,
        Karl-Schwarzschild-Strasse 1, 85741 Garching, Germany \\
        $^5$Kavli Institute for Cosmology and Institute of Astronomy, 
        Madingley Rd, Cambridge CB3 0HA, UK }

\begin{document}
\date{Accepted ???. Received ??? in original form ???}

\maketitle
\label{firstpage}

\begin{abstract}
We employ cosmological hydrodynamical simulations to investigate
the effects of  AGN feedback on the formation of massive galaxies 
with present-day stellar masses of $\Mstel = 8.8 \times 10^{10} - 6.0 
\times 10^{11} \Msun$. Using smoothed particle hydrodynamics 
simulations with a pressure-entropy formulation that allows an 
improved treatment of contact discontinuities and fluid mixing, we run 
three sets of simulations of 20 halos with different AGN feedback 
models: (1) no feedback, (2) thermal feedback, and (3) mechanical 
and radiation feedback. We assume that seed black holes are 
present at early cosmic epochs at the centre of emerging dark matter 
halos and trace their mass growth via gas accretion and mergers with
other black holes. Both feedback models successfully recover the 
observed $\Mbh - \sigmastar$ relation and black hole-to-stellar mass 
ratio for simulated central early-type galaxies. The baryonic 
conversion efficiencies are reduced by a factor of two compared to 
models without any AGN feedback at all halo masses. However, 
massive galaxies simulated with thermal AGN feedback show a factor 
of $\sim10-100$ higher X-ray luminosities than observed. The 
mechanical/radiation feedback model reproduces the observed 
correlation between X-ray luminosities and velocity dispersion, e.g. 
for galaxies with $\sigmastar=200$ $\kms$, the X-ray luminosity is 
reduced from $10^{42}$ $\ergs$ to $10^{40}$ $\ergs$. It also 
efficiently suppresses late time star formation, reducing the specific 
star formation rate from $10^{-10.5}$ $\yr$ to $10^{-14}$ $\yr$ on 
average and resulting in quiescent galaxies since z=2, whereas the 
thermal feedback model shows higher late time in-situ star formation 
rates than observed.
\end{abstract}


\section{Introduction}
There is solid observational evidence that most massive galaxies 
harbor massive black holes in their centres 
\citep{1995ARA&A..33..581K}, the masses of which are 
correlated with the properties of their host galaxies, including bulge 
mass \citep{1998AJ....115.2285M,2003ApJ...589L..21M}, velocity 
dispersion \citep{2000ApJ...539L..13G,2000ApJ...539L...9F,
2002ApJ...574..740T}, globular cluster systems 
\citep{2010ApJ...720..516B}, and many others
\citep[see][and references therein]{2013ARA&A..51..511K}.
It was proposed that the energy released by the accretion mechanism 
onto black holes is sufficient enough to unbind galactic interstellar 
medium and trigger strong outflows \citep{1998A&A...331L...1S}.
This AGN feedback mechanism is confirmed by many observations
including the broad absorption lines in the spectra of quasars, the
detection of  X-ray cavities and radio jets
 \citep[see][and references therein]{2012ARA&A..50..455F}.

Without the inclusion of AGN feedback traditional numerical 
cosmological simulations of the formation of massive early-type 
galaxies suffer from the `overcooling' problem: over a Hubble time 
too much cold gas is able to cool at the centres of the galaxy halos 
\citep{1993ApJ...412..455K}. This leads to overly massive and 
actively star-forming ellipticals at present, in contradiction with 
observations \citep[e.g.][]{2003MNRAS.341...33K,
2008ApJ...688..770F} and basic predictions of \LCDM structure 
formation \citep[e.g.][]{2010ApJ...710..903M}. Many semi-analytical 
models of galaxy formation, extended by various models for AGN 
feedback, have demonstrated that this process is relevant for 
reproducing the bright end of the galaxy luminosity function
\citep{2006MNRAS.370..645B,2006MNRAS.365...11C,
2007MNRAS.375....2D,2008MNRAS.391..481S,
2011MNRAS.413..101G}. Dynamically and spatially better resolved 
direct hydrodynamical cosmological simulations including AGN 
feedback confirm that this mechanism is able to self-regulate the 
stellar baryon content in massive halos resulting in more realistic 
galaxy masses and colours, and that it can reproduce the relations 
between black hole mass and bulge mass ($M_{\rm BH}-M_{\rm b}$) 
and stellar velocity dispersion $M_{\rm BH}-\sigmastar$
\citep[e.g.][]{2007MNRAS.380..877S,
2011MNRAS.413.1158B,2011MNRAS.414..195T,
2012MNRAS.422.3081M,2013MNRAS.436.3031V}.

Many previous AGN feedback models include only a subset of the 
known and observed feedback processes. Since the importance of 
AGN-driven mass and momentum outflows in limiting the infall onto 
the black hole has been emphasized 
\citep[see][]{2005ApJ...618..569M,2010ApJ...722..642O,
2010MNRAS.406L..55D,2011MNRAS.412.1341D}, 
in previous works \citep{2012ApJ...754..125C,
2014MNRAS.442..440C} we proposed a mechanical and radiative 
AGN feedback together with the detailed treatment of radiative 
heating, radiation pressure, and the Eddington force from black holes 
and investigated the effects of the new AGN feedback model. It is 
shown that the new feedback model can regulate the black hole 
growth, and successfully reproduce the observed and 
$\Mbh - \sigmastar$ relation in a way similar to the successful 
thermal feedback approach \citep{2005MNRAS.361..776S}. But as 
pointed out in \cite{2010MNRAS.406..822M}, the X-ray output is 
sensitive to AGN feedback. The commonly adopted thermal feedback 
model where all the feedback energy is distributed as local thermal 
heating produces a factor of  $\sim$100 higher thermal X-ray 
luminosity than expected for a given stellar mass of the galaxy while 
the new approach reproduces the observed $L_X - \sigmastar$ 
\citep{2012ApJ...754..125C,2014MNRAS.442..440C}. The physical 
reason for this is simple: momentum input from AGN winds and 
radiation efficiently drives gas out into the surrounding halo, reducing 
the central density and thermal X-ray luminosity. The two treatments - 
thermal vs mechanical - put in the same total energy for a given 
accretion rate and given efficiency, but putting some fraction of the 
energy into mechanical rather than thermal increases the 
effectiveness in driving gas out of the galaxy. The same result was 
found in supernova feedback \citep{2014arXiv1410.3822S}. The 
mechanical feedback model also shows large fluctuations in both 
radiant and wind outputs, in agreement with observations
\citep[see also][]{2012MNRAS.420.2221D}.

In a next step we now aim at characterizing the evolution of galaxies 
and black holes with mechanical and radiation AGN feedback in a
full cosmological context. Even though numerical limitations impose 
inevitable approximations in the form of sub-resolution models, 
cosmological simulations can provide a more complete view on the 
evolution of galaxies and their central black holes, starting from initial 
conditions consistent with the early universe as specified by the 
WMAP \citep{2003ApJS..148..175S} and Planck satellites 
\citep{2013arXiv1303.5062P}. In order to statistically confront 
observations, we need a cosmological sample of massive galaxies 
and black holes that covers a wide range of environments and 
redshifts.

In this paper, we investigate the effects of mechanical and radiation
AGN feedback on the cosmological formation of individual galaxies 
with three sets of cosmological, hydrodynamical zoom simulations of 
20 halos in the mass range $2.3 \times 10^{12} \Msun$ $ \lesssim 
M_{\mathrm{vir}} \lesssim 1.4 \times 10^{13} \Msun$.
This extends the recent studies by \cite{2010ApJ...725.2312O,
2012ApJ...744...63O}, on which our initial conditions are based, but 
did not contain AGN feedback. In the three simulation sets, we add 
the commonly adopted thermal AGN feedback and our mechanical 
and radiation AGN feedback separately to investigate their impact on 
black hole growth, the conversion of gas into stars, star formation, and 
thermal X-ray luminosity of hot gas.
  
The paper is organized as follows. Section~\ref{model_cosmo} 
provides an introduction to our simulations and the adopted AGN 
feedback model. In section~\ref{result_cosmo}, we discuss the effect 
of AGN feedback on black hole growth, black hole and galaxy scaling 
relations, X-ray luminosities of hot gas, baryonic stellar and gas mass 
fraction, star formation rates, galaxy sizes and velocity dispersions. A 
final summary and discussion of this work is given in 
section~\ref{summary_cosmo}.

\section{Simulations}\label{model_cosmo}
\subsection{Numerical code}
The simulations presented in this paper were performed with a 
modified version of the parallel smoothed particle hydrodynamics
(SPH) code GADGET-3 \citep{2005MNRAS.364.1105S}. Although 
SPH has many advantages, such as the exact conservation of 
physical properties and the adaptive resolution, recent studies have 
shown that standard SPH has severe difficulties in modeling fluid 
mixing owing to spurious surface tension at contact discontinuities 
\citep[e.g.][]{2007MNRAS.380..963A}. To avoid the numerical 
artifacts we have used \textsc{SPHGal} a modified version of 
GADGET with improved accuracy. Details of the code are presented 
in \citet{2014MNRAS.443.1173H}, in the following we give a brief 
overview. 

The code includes a density-independent SPH formulation by 
choosing a different volume element \citep{2001MNRAS.323..743R,
2013ApJ...768...44S}. Specifically, we employ the pressure-entropy 
formulation as described in \cite{2013MNRAS.428.2840H}. 
Furthermore, we improve the force accuracy of our method by 
increasing the particle number in the kernel. To avoid the pairing 
instability we adopt the Wendland $C^4$ kernel with 200 
neighboring particles \citep{2012MNRAS.425.1068D}. To further 
improve over standard SPH we use the improved artificial viscosity 
implementation presented by \cite{2010MNRAS.408..669C}. In this 
scheme each SPH particle has a variable viscosity coefficient which 
only increases when a converging flow is detected and afterwards 
decays to a minimum in a few sound-crossing times such that 
unwanted viscosity away from shocks is suppressed. To detect 
shocks in advance, the time derivative of the velocity divergence is 
used. We further include  an artificial thermal conductivity following 
\cite{2012MNRAS.422.3037R} which smoothes the internal energy 
while explicitly conserving it within the kernel. While this seems to be 
redundant in the pressure-entropy formulation, in strong shocks the 
entropy jumps can be very high resulting in very noisy pressure 
estimates. Artificial diffusion is reduced away from entropy 
discontinuities with a switch similar to the one used for artificial 
viscosity. Finally we employ a time-step limiter  following 
\citet{2009ApJ...697L..99S,2012MNRAS.419..465D} to ensure that 
neighboring particles have similar time-step (within a factor of 4), 
such that ambient particles do not remain inactive when a shock is 
approaching. The performance of the new SPH schemes in the test 
problems is discussed in \citet{2014MNRAS.443.1173H}.

For the star formation and feedback prescription we use the 
self-regulated supernova feedback model of 
\citet{2003MNRAS.339..289S}. This model treats the interstellar 
medium (ISM) as a two-phase medium \citep{1977ApJ...218..148M} 
where clouds of cold gas are embedded in the hot gas phase at 
pressure equilibrium. Stars are allowed to form out of the cold gas 
phase if the local density exceeds a threshold value ($n > n_{th} = 
0.205 \cm^{-3}$) which is calculated self-consistently in a way that 
the equation of state is continuous at the onset of star formation. 
Finally the simulations include a cooling prescription for a primordial 
composition of hydrogen and helium and a redshift-dependent UV 
background radiation field with a modified \citet{1996ApJ...461...20H} 
spectrum.

\subsection{Simulation setup}
\begin{table}
   \begin{center}
   \caption{Full list of parameters of the two AGN models }
    {
   \begin{tabular}{c|c|c}\hline\hline   \label{tab:parameters}
parameters & ThAGN$^{a}$ & MrAGN$$ \\
  \hline
$M_{\rm BH seed }$ &  $10^5 \Msunh$ & $10^6 \Msunh$ \cr
$M_{\rm halo}$ & $5 \times 10^{10} \Msunh$ & $10^{11} \Msunh$\cr
$\alpha$ & 100 & 1 \cr
$\epsw$ & 0.005 & 0.002 \cr
  \hline\hline
   \end{tabular}}
   \end{center}
   \begin{flushleft}
    $^{a}$ Note that we use the standard mass accretion prescription 
    and parameters used in the previous works 
    \citep[e.g.][]{2007MNRAS.380..877S}, which produce a broad 
    agreement with observational constraints.
   \end{flushleft}  
 \end{table}
 
We use the cosmological `zoom-in' initial conditions which are 
described in detail in \citet{2010ApJ...725.2312O}. The halos are 
picked from a dark matter only simulation using a flat cosmology with 
parameters obtained from WMAP3 \citep{2007ApJS..170..377S}: 
$h=0.72, \; \Omega_{\mathrm{b}}=0.044, \; 
\Omega_{\mathrm{dm}}=0.216, \;
\Omega_{\Lambda}=0.74, \; \sigma_8=0.77 $ and an initial slope of 
the power spectrum of $\mathrm{n_s}=0.95$. From redshift zero we 
trace back in time all particles close to the halos of interest at any 
given snapshot. Those particles are then replaced with 
high-resolution gas and dark matter particles.  The original dark 
matter particles are merged to reduce the particle count and the 
simulation time depending on their distance to the re-simulated halo.  
The new high resolution zoom-in initial conditions are evolved from 
redshift z=43 to the present day.

The simulated halo masses cover the range   $2.3 \times 10^{12}
\Msun$ $ \lesssim M_{\mathrm{vir}} \lesssim 1.4 \times
10^{13} \Msun$ and the central galaxy masses are between
$8.8 \times 10^{10} \Msun \lesssim M_* \lesssim 6.0 \times
10^{11} \Msun$ at $z=0$. The masses for the gas and star particles 
are $m_{*,gas}=4.2 \times 10^{6} \Msunh$ (Note that we spawn one 
star particle per gas particle), whereas the dark matter particles have 
a mass of $m_{\mathrm{dm}} = 2.5 \times 10^{7} \Msunh$. The 
comoving gravitational softening lengths used are 
$\epsilon_{\mathrm{gas,star}} = 400 \rm \pch $ for the gas and star 
particles and $\epsilon_{\mathrm{halo}} = 890 \rm \pch$ for the dark 
matter scaled with the square root of the mass ratio 
\citep{2001MNRAS.324..273D}. In the following we present the 
results for 20 galaxies with masses larger than 
$M_* \approx 8.8 \times 10^{10} \Msun$ for direct comparison with 
observations.  These galaxies are well resolved with 
$\approx 1.2 - 6.8 \times 10^5$ particles within the virial radius. Using 
the above simulation parameters for zoom simulations has been 
shown to result in galaxies with reasonable present-day properties 
\citep{2007ApJ...658..710N,2009ApJ...697L..38J,
2009ApJ...699L.178N,2010ApJ...725.2312O}. However, 
\citet{2010ApJ...725.2312O} shows that the fraction of available 
baryons converted into stars, $f_*$, for galaxies in this mass range is 
typically 2 times higher than estimates from models that are 
constructed by matching observed luminosity functions to simulated 
halo mass functions \citep[and references therein]
{2010ApJ...710..903M,2010MNRAS.404.1111G,
2010ApJ...717..379B,2013MNRAS.428.3121M}.

In order to study the effects of numerical resolution, we run the high 
resolution simulation for a high-mass halo with $M_{\mathrm{vir}} 
\sim 5.0 \times 10^{12} \Msun$. The high resolution initial condition 
has twice better spatial resolution and eight times better mass
resolution than our fiducial resolution 
(i.e. $\epsilon_{\mathrm{gas,star}} = 200 \rm \pch$ and 
$m_{*,gas}=5.3 \times 10^{5} \Msunh$).

\subsection{Black hole formation and growth}\label{sec:bh_acc}
In our cosmological simulations of structure formation black holes are 
modeled as collisionless sink particles which form in newly emerging 
dark matter halos. We assume that black holes are seeded such that 
any halo above a certain threshold mass contains one black hole at 
its centre. We identify haloes on the fly during a simulation by calling 
a friends-of-friends (FOF) algorithm at regular intervals. All halos with 
a mass larger than $M_{\rm halo } =10^{11} \Msunh$ are provided 
one black hole with mass of $M_{\rm BH seed } =10^6 \Msunh$ if 
they do not contain any black hole already. The black hole seed 
mass and halo threshold mass are chosen to roughly follow the 
Magorrian relation \citep{1998AJ....115.2285M} and the chosen 
seed mass is small enough that it only makes a negligible 
contribution to the mass of the final black hole. Growth of 
supermassive black holes has been previously modeled with full 
cosmological hydrodynamical simulations with the chosen black hole 
seed mass ranges from $10^5 \Msunh$ to $10^6 \Msunh$, and a 
threshold mass for halo from $10^{9} \Msunh$ to $10^{11} \Msunh$, 
and it is shown that these seeds evolve into a population of 
supermassive black holes with masses and accretion luminosities in 
line with observational estimates by $z \sim 6$ 
\citep{2009MNRAS.400..100S,2012ApJ...745L..29D,
2012MNRAS.423.2397K,2012MNRAS.424.1892D}.
It has also been shown that the AGN behavior is not particularly 
sensitive to the seed mass \citep{2006ApJ...639..700H}.

The mass of black holes is assumed to grow via two channels:
mergers with other black holes and accretion of gas. Two black hole 
particles are allowed to merge if they fall within their local SPH 
smoothing lengths and if their relative velocities are smaller than the 
local sound speed. The rate of the gas infall onto the black hole is 
estimated with a Bondi-Hoyle-Lyttleton parameterization 
\citep{1939PCPS...34..405H,1944MNRAS.104..273B,
1952MNRAS.112..195B} with an alternative averaging method 
introduced in \cite{2012ApJ...754..125C}. For gas with the density 
$\rho$, a sound speed $c_{\rm s}$ and the velocity relative to the 
black hole $v$, the mass accretion rate onto the central region is:
\begin{equation}
\dot{M}_{\rm{inf}}= 
\left\langle \frac{4 \pi  G^{2} M_{\rm BH}^{2} \rho }
                            {(c_{\rm s}^2+ v^{2})^{3/2}} \right \rangle,
\label{bondi_AA}
\end{equation}
where angle brackets denote the averaging over the SPH kernel. 
This method for the calculation of the black hole mass does the 
calculation in both time and space on an individual particle basis 
and then averages the results over the neighboring 64 particles in 
order to reduce the dependency on the number of SPH particles.

To avoid the unphysical accretion of unbound gas from outside the
Bondi radius which occurs in some treatments of this problem we 
statistically limit the accretion of mass to the gas within the Bondi 
radius. Since the mass distribution of each gas particle is smoothed 
with the kernel size, we allow for the full accretion rate only if the total 
volume of a gas particle resides within the Bondi radius. Otherwise, 
we reduce the probability of being absorbed by the black hole (soft 
Bondi radius criterion, see \cite{2012ApJ...754..125C}). To account 
for the time that it takes a particle at radius $r_j$ to be accreted, we 
include the free-fall modification to the accretion probability with an 
extra factor of
\beq
p_{j,\rm ff} = \frac{ \frac{1}{\tau_j}}
{ \frac{1}{N_{\rm sph}} \sum\limits_{j=1}^{N_{\rm sph}} \frac{1}{\tau_j}},
\label{p_ff}
\eeq
where  $\tau_j = {r_j} / (c_{\rm s,\it j}^2+v_{j}^{2})^{1/2}$ is the free fall 
time and $N_{\rm sph}$ denotes the typical number of smoothing 
neighboring gas particles of the black hole. For a full description of 
the soft Bondi radius criterion and the free-fall modification, see 
Figure~1 and section 2.4 of \cite{2012ApJ...754..125C}.  

We do not employ an additional factor `$\alpha$' with regard to the 
accretion rate. This has been utilized to overcome resolution 
problems and better match reality, but adopting it increases the 
uncertainty of the numerical treatment 
\citep{2009MNRAS.398...53B}.

\subsection{Mechanical and radiation black hole feedback}
Motivated by observations of broad absorption line winds, which 
convey energy, mass and momentum into the surrounding gas with 
velocity $\sim 10,000$ $\kms$ outflows corresponding to a typical 
broad line wind velocity \citep{2003ARA&A..41..117C,
2009ApJ...706..525M,2010ApJ...709..611D}, we included these 
observed AGN winds in our numerical treatment following 
\citet{2010ApJ...722..642O}. In our model, the AGN winds carry a 
mass given by:
\beq
\Mdotoutf= \Mdotinf - \Mdotacc, \label{eq:Mdot}
\eeq
where $\Mdotoutf$, $\Mdotinf$ and $\Mdotacc$ respectively denote 
the outflowing/inflowing mass rate and the mass rate actually 
accreted onto the black hole. For simplicity we assume that the wind 
is launched at a fixed speed $\vw =10,000$~$\kms$. Then a 
momentum flux carried by the wind is given as,
\beq
\pdot = \Mdotoutf \vw, \label{eq:pdot}
\eeq
and the kinetic energy rate of the outflow is given as,
\begin{subequations}
\begin{eqnarray}
\Edotw & \equiv & \epsw \Mdotacc c^2, \label{eq:edotw1} \\
      &=& \frac{1}{2} \Mdotoutf \vw^2,\label{eq:edotw2} 
\end{eqnarray}
\end{subequations}
where $\epsw$ denotes the feedback efficiency.  We can define the 
dimensionless quantity $\psi$, the ratio of the mass outflow rate to 
the accreted rate as,
\beq
\psi \equiv 2 \epsw c^2 / \vw^2=\Mdotoutf / \Mdotacc,
\label{eq:psi}
\eeq
and we can rewrite the equation for the black hole accretion rate as,
\beq
\Mdotacc = \Mdotinf \frac{1}{1+\psi}\label{eq:Mdot_sol}.
\eeq
As discussed in \citet{2010ApJ...722..642O,2012ApJ...754..125C} in 
the presence of significant AGN winds, not all of the mass entering 
the central region $\Mdotinf$ actually reaches the black hole. For 
example, with the feedback efficiency typically adopted in the 
literature, $\epsw = 0.005$, and with the fixed wind velocity 
$\vw = 10,000$~$\kms$, equation (6) and (7) which are based on 
mass and energy conservation indicate that only 10 percent of the 
inflowing mass is actually accreted onto the black hole while 90 
percent is ejected in a wind.

We calculate the dimensionless quantity $\psi$ for the given 
feedback efficiency $\epsw$ and wind velocity $\vw$, and 
stochastically select the wind particles from all gas particle attracted 
into the central zone by the black hole keeping the fraction of wind 
particles to the total inflowing particles as $\psi/(1+\psi)$. To deposit 
the wind mass and momentum, we give kicks to the gas particles 
selected following the stochastic approach. We set the direction of the 
wind to be parallel or anti-parallel to the direction of angular 
momentum of each gas particle, if the central black holes are 
surrounded by a gas disc this procedure results in a wind 
perpendicular to the disc plane \citep{2004ApJ...616..688P}. The 
emitted wind particles share their momentum with two other nearby 
gas particles to reproduce the shock heated momentum-driven flows. 
They have the same velocity increment, $\Delta v \sim 10,000/3$ 
$\kms$, conserving the momentum. But sharing momentum with other 
particles via inelastic collisions decreases the total kinetic energy 
increment while preserving momentum. We deposit the residual 
energy into these three particles in thermal form so that the total 
energy is conserved. Having momentum sharing with two nearby gas 
particles gives 2:1 divisions into thermal and kinetic energies so the 
wind particles can reach very high temperatures. This ratio is similar 
to that in the standard Sedov-Taylor self-similar blast wave.

\begin{figure}
\epsfig{file=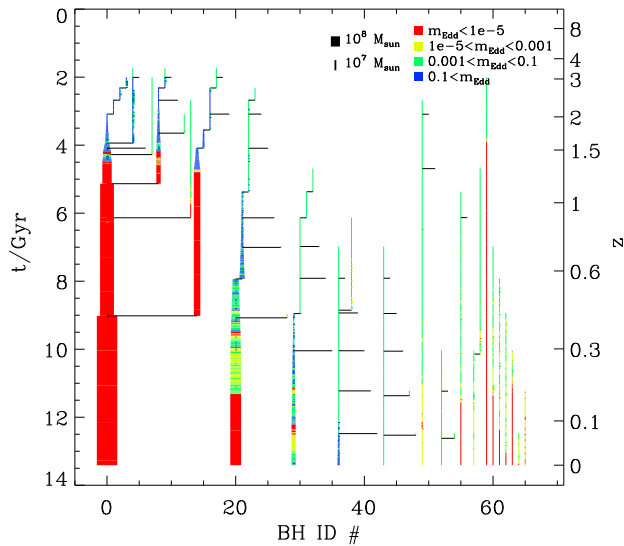,width=\columnwidth}
\caption{A mass assembly tree of black holes in a sample zoom-in
region with mechanical and radiation AGN feedback. Each line 
represents a black hole and the thickness of the lines is scaled by the 
mass of the black hole. Mass accretion rate onto the black hole is 
colour-coded in Eddington units, as black hole mass accretion with 
Eddington ratio $\mEdd<10^{-5}$ shown in red, 
$10^{-5}<\mEdd<10^{-3}$ in yellow, $10^{-3}<\mEdd<10^{-1}$ in 
green, and $10^{-1}<\mEdd$ in blue. In total 66 black holes are 
seeded  in this zoom-in region, after black hole mergers 16 black 
holes survive to $z = 0$. The leftmost branch of the merger tree 
shows the central black hole with a final mass of 
$\sim 2.2 \times 10^{8} \Msunh$ which resides in a main halo with a 
mass of $\sim 5 \times 10^{12} \Msunh$. The leftÐ-right positioning of 
the progenitors is schematic, and  is not connected to the position 
within the dark matter halo. 
\label{fig:bhtree}}
\end{figure}

In addition to the mechanical feedback described above, X-ray 
radiation from the accreting black hole can be coupled to the 
surrounding gas according to an approximation described in
\cite{2005MNRAS.358..168S}, as in \cite{2010ApJ...717..708C,
2011ApJ...737...26N,2012ApJ...754..125C,2014MNRAS.442..440C}; 
The luminosity flux from the multiple black holes is calculated at the 
position of each gas particle, and the flux is converted to the net 
volume heating rate $\dot{E}$ by adopting the 
\cite{2005MNRAS.358..168S} formulae that include Compton heating 
and photoionization heating. Note that Equation~\ref{eq:Mdot_sol}, 
not Equation~\ref{bondi_AA}, determines the AGN luminosity flux and 
thus the magnitude of the radiation feedback. We also include the 
electromagnetic momentum, the radiation pressure from the X-ray flux 
from the black hole by adding a momentum per unit time of 
$\dot{p}=\dot{E}/c$. The added force is directed radially away from 
the black holes.

The flux from UV bump in AGN spectrum dominates for the 
momentum driven winds as shown in high resolution hydrodynamical
simulations \citep[e.g.][]{2000ApJ...543..686P}. The region where 
this driving occurs is very close to the quasar so we include this in 
sub-grid modeling, as in ``mechanical feedback". And the radiation 
heating and its associated radiation pressure are from the 
moderately hard X-ray region ($\sim 50$ keV), which dominates 
the heating process. Thus the momentum in the AGN UV emission 
is included indirectly by the absorption of the broad absorption line 
wind that has been accelerated  by the metal line trapping 
\citep{2000ApJ...543..686P}, and the momentum in the X-ray bump 
is partially allowed for in so far as the X-rays have been absorbed 
following a standard atomic physics treatment.

Finally, instead of limiting the maximum accretion rate to the 
Eddington rate, we compute the Eddington force acting on the 
surrounding gas particles, directed radially away from the black hole 
as described in \citet{2012ApJ...754..125C} and allow this force to 
act on the gas flow through the hydrodynamic equations. Naturally it 
reduces the inflow and increases the outflow but accretion exceeding 
the Eddington rates can occasionally occur.

In order to study the effects of different AGN feedback models, we 
investigate the influence on galaxy separately by running the full set 
of simulations with three different model:

{\bf (1) NoAGN:} No black hole and no AGN feedback. This model is 
comparable to the results of \cite{2010ApJ...725.2312O,
2012ApJ...744...63O}, but note that we use an alternative formulation 
of SPH designed to treat contact discontinuities more accurately, and 
include an improved artificial viscosity and an energy diffusion
implementation. 

{\bf (2) ThAGN:} with black hole and the classical thermal AGN 
feedback. Note that we use the standard mass accretion prescription 
and parameters adopted and studied in the previous works of 
\cite{2007MNRAS.380..877S,2009MNRAS.400..100S}, i.e. $10^5 
\Msunh$ for black hole seed mass, $5 \times 10^{10} \Msunh$ for 
threshold halo mass of the black hole seeding, and $\epsw=0.005$
for AGN feedback efficiency which produce a broad agreement with 
observational constraints.

{\bf (3) MrAGN:} with black hole and mechanical and radiation AGN 
feedback. This model includes the modified black hole mass 
accretion described in Section~\ref{sec:bh_acc}. We use lower 
feedback efficiency  $\epsw=0.002$ for this model which was 
constrained based on our previous small scale simulations 
\citep{2014MNRAS.442..440C}. The adopted black hole and AGN 
feedback related parameters are shown in Table~\ref{tab:parameters}.

We only consider the central galaxies within the simulated halos and 
their black holes in this study.

\section{Cosmological simulations of AGN feedback in
individual galaxy halos}\label{result_cosmo}
\subsection{Black hole growth}
\begin{figure}
\epsfig{file=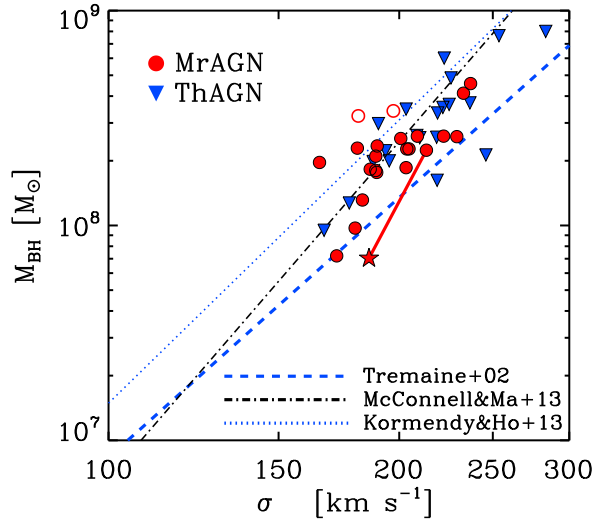,width=\columnwidth}
\caption{The black hole mass-stellar velocity dispersion 
($\Mbh - \sigmastar$) relation at $z=0$ of the two AGN feedback 
models: thermal feedback (ThAGN, blue upside down triangles) and 
momentum and radiative feedback (MrAGN, red circles). The red star 
symbol shows the higher resolution run for one halo MrAGN model 
and red solid line connects the corresponding lower resolution run.
The momentum feedback runs without radiative feedback are shown 
in open red circles for two halos. The black/blue dotted lines show
the observed relation of the early type galaxies from McConnell \& 
Ma (2013) and Kormendy \& Ho (2013) respectively.
\label{fig:msigma}}
\end{figure}

\begin{figure}
\epsfig{file=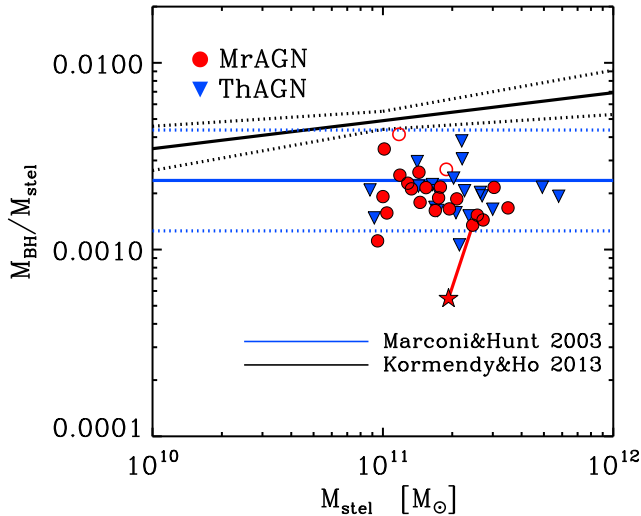,width=\columnwidth}
\caption{The black hole mass-to-stellar mass ratio at $z=0$ is plotted 
against stellar mass for two AGN feedback models: thermal feedback 
(ThAGN, blue upside down triangles) and momentum and radiative 
feedback (MrAGN, red circles). The high resolution run with 
momentum feedback is shown in star symbol with the red solid line 
that connects the corresponding low resolution run. The momentum 
feedback runs without radiative feedback are shown in open red 
circles for two halos. The blue and black solid lines show the 
observed relation from Marconi \& Hunt (2003) and Kormendy \& Ho 
(2013) respectively with an intrinsic dispersion shown in dotted lines.
\label{fig:mbh_mstel}}
\end{figure}

In order to show how individual black holes grow in our simulations, 
we show in Figure~\ref{fig:bhtree} an example of the mass assembly 
history of black holes in a sample zoom-in region which forms a main 
halo with a mass of $\sim 5 \times 10^{12} \Msunh$ at $z=0$ with the 
MrAGN model. Each line represents a black hole and the thickness 
of the lines is scaled by the mass of the black hole. Mass accretion 
onto the black hole is colour-coded in Eddington units, as black hole 
mass accretion with Eddington ratio $\mEdd<10^{-5}$ shown in red, 
$10^{-5}<\mEdd<10^{-3}$ in yellow, $10^{-3}<\mEdd<10^{-1}$ in 
green, and $10^{-1}<\mEdd$ in blue. In total 66 black holes are 
seeded, after black hole mergers only 16 black holes survive to 
$z=0$. The left--right positioning of the progenitors is schematic, and 
has no relevance to spatial positions of galaxies within the dark 
matter halo. The leftmost branch of the merger tree shows the 
central black hole with a final mass of $\sim 2.3 \times 10^{8} \Msun$.
At $z = 0$, the most massive black hole is growing less rapidly than 
less massive black holes.  At higher redshift, this trend is greatly 
diminished, and at z = 2 it is reversed so that the most massive black 
holes have the highest fractional accretion rate. This phenomenology 
broadly agrees with the observed `cosmic downsizing' of AGN, i.e. 
the most massive and luminous AGN were most numerous at redshift 
$z\sim 2-2.5$, less luminous AGN peaked at successively lower 
redshifts with the least luminous peaking at around $z \sim 1$ in 
X-rays \citep[e.g.][]{2003ApJ...598..886U}, and in optical and other 
bands \citep[e.g.][]{2011MNRAS.416.1900R}.

\begin{figure*}
\epsfig{file=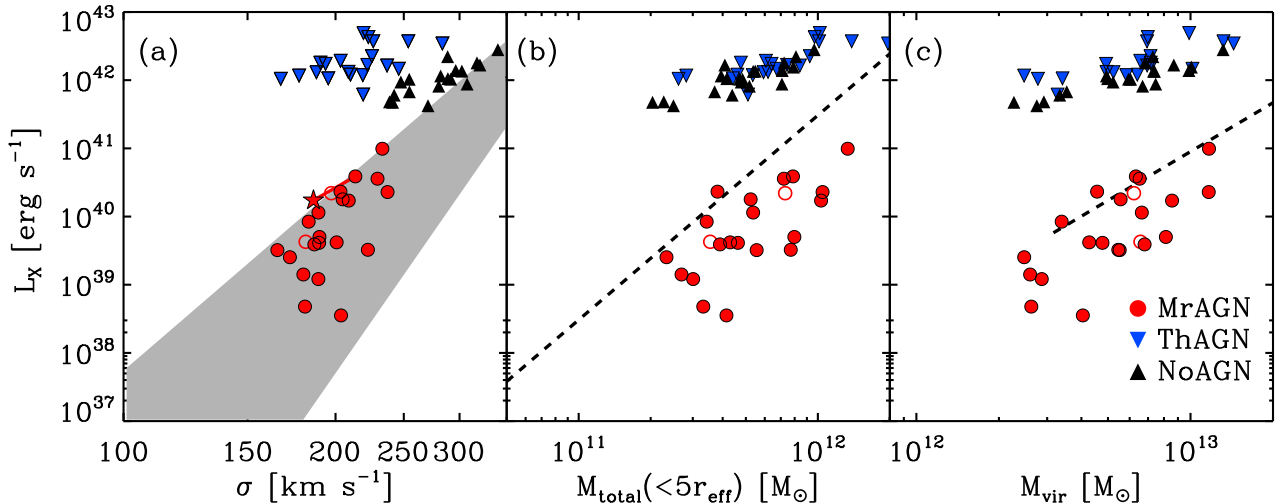,width=\textwidth}
\caption{
X-ray luminosity versus (a) stellar velocity $\sigmastar$, (b) total 
mass within $5 \re$, and (c) virial mass $\Mvir$ of the simulated 
galaxies at $z=0$ for the model without AGN feedback (NoAGN, 
black triangles), thermal feedback (ThAGN, blue upside down 
triangles), and momentum and radiative feedback (MrAGN, red 
circles). The high resolution run with momentum feedback is shown 
in star symbol with the red solid line that connects the corresponding
low resolution run. The momentum feedback runs without radiative 
feedback are shown in open red circles for two halos. Observed 
relations for the normal early-type galaxies are from 
(a) \citet{2011ApJ...729...12B}, (b) \citet{2013ApJ...776..116K}, and 
(c) \citet{2006ApJ...652L..17M} respectively. The simulations without 
AGN feedback and with thermal AGN feedback show systematically 
higher X-ray luminosities than observed.  
\label{fig:xray_cosmo}}
\end{figure*}

We now study how the black hole masses relate to their host 
galaxies. For this purpose, we determine the central galaxy 
properties using a spherical over-density criterion at $z=0$. 
Following previous studies \citep[e.g.][]{2012ApJ...744...63O},
we obtain a radius where the spherical over-density drops below 
200 times the critical density of the universe for a virial radius of the 
halo, $\rvir$, and the virial mass $\Mvir$ is defined by total mass 
therein. We then determine the effective radius of the galaxy $\re$ by 
getting the mean values of the half-mass radii of all stars within 
$10 \%$ of the virial radius $\r10$ $= 0.1 \times  \rvir$ projected 
along the three principal axes of the main stellar body. The 
line-of-sight velocity dispersions $\sigmastar$ have been calculated 
within $0.5 \times \re$ along the three principal axes and then 
averaged.

In Figure~\ref{fig:msigma}, we show the relation between black hole 
mass and stellar velocity dispersion $\sigmastar$ for the two 
feedback models. Blue upside down triangles are for the simulations 
with the classical thermal AGN feedback, ThAGN, while red circles 
are for the simulations with the mechanical and radiation AGN 
feedback, MrAGN. Note that two feedback models are tested for the 
same set of 20 halos. We overplot the observed $\Mbh - \sigmastar$ 
relations as presented in recent papers by \cite{2013ApJ...764..184M} 
(black) and \cite{2013ARA&amp;A..51..511K} (blue) respectively. We 
find a good agreement with the observations for both AGN feedback
models as found in our previous study on the bulge dominated 
merger remnants of the two disc galaxies 
\citep{2014MNRAS.442..440C}. In previous works on black hole 
growth with the classical thermal AGN feedback 
\citep[e.g.][]{2007MNRAS.380..877S,2008ApJ...676...33D}, 
a measurement of the $\Mbh - \sigmastar$ relation for the simulated
galaxies was also found to successfully produce the observed black 
hole - bulge relations. Note that we use the parameters adopted in 
\citet{2007MNRAS.380..877S}, but we use the modified version of 
GADGET-3 code with an alternative formulation of SPH, an updated 
artificial viscosity and an energy diffusion implementation. The 
modifications of the code we include in this study do not change their 
main findings on the black hole mass and bulge relation in general.

In order to study the relative effect of radiation feedback, we perform 
the control runs for two halos including only the mechanical wind 
feedback. The results from control run without the radiative feedback 
are shown in open red circles. The radiative feedback shows 
moderate impact on the growth of black hole, but the differences are 
within the observed scatters.

We also perform a resolution study for our primary example with a 
halo mass of $5 \times 10^{12} \Msunh$ with MrAGN feedback. The 
high resolution run with twice better spatial resolution and with eight 
times better mass resolution is shown in red star symbol with the red 
solid line connecting corresponding fiducial resolution run. The 
resolution has a strong effect both on the velocity dispersion (the final 
stellar mass) of the galaxy and on the final black hole mass. In higher 
resolution the final black hole mass is lower by a factor of 3, and this 
indicates that with resolution increased the black hole feedback
becomes slightly more efficient. As the differences induced by 
increasing resolution are within the observed uncertainties on 
M-sigma relation, we restrict our main study to our fiducial resolution 
in this paper. The convergence will be further studied and discussed 
in later papers.

Figure~\ref{fig:mbh_mstel} shows the black hole--to--stellar mass 
ratio as a function of  galaxy stellar mass for the simulated two 
feedback models: blue upside down triangles are for ThAGN models 
and red circles are for MrAGN models. The observed relations from 
\cite{2003ApJ...589L..21M} and  \cite{2013ARA&amp;A..51..511K} 
are shown in blue lines and black lines respectively. The simulated 
relations at $z=0$ for both AGN feedback models match the locally 
observed canonical black hole--to--bulge mass ratio $0.23 \%$ from 
\cite{2003ApJ...589L..21M} well, but the simulated ratios are $\sim 2$ 
times lower than the most recent relation found by 
\cite{2013ARA&amp;A..51..511K} who strictly limit their samples and 
omit all pseudo bulges and merging galaxies. Since we have not 
made such a restriction for measuring stellar mass in our simulated 
galaxies, our result would be more appropriately compared to the 
previously found values of the canonical black hole--to--bulge mass 
ratios.

\subsection{Impact of AGN feedback on X-ray luminosity}
The thermal AGN feedback prescription adopted in many previous 
AGN feedback studies hides the small-scale physics such as 
radiation-driven winds and shocks under a simple parameterization 
of local input, and only treats the resulting effects of feedback. As a 
result, the gas is simply heated near the black hole and produces a 
very large thermal X-ray luminosity. In the mechanical AGN feedback 
model, however, the X-ray luminosity is lower, because the 
outflowing gas carries off significant kinetic energy, which tends to 
drive gas away from the galaxy, reduce its density and significantly 
reduce its thermal X-ray output. Hence comparing our simulations 
with the observations of X-ray scaling relations of nearby early type 
galaxies \citep[e.g.][]{1985ApJ...293..102F,1985ApJ...296..447T, 
2003ARA&amp;A..41..191M,2006ApJ...646..899H} should give us a
clean discriminant between the AGN feedback models.

We calculate the X-ray luminosity due to bremsstrahlung radiation as 
well as line emission from all relevant species in {\it Chandra} bands 
(0.3-8 keV) for the simulated central elliptical galaxies  using the tools 
developed in \cite{2014MNRAS.442..440C}. In order to exclude the 
star forming gas in the obscured central region of the galaxy, we only 
include the hot and diffuse gas within the virial radius with a 
temperature and density cut of $T \geq 10^6$ K, and 
$\rho \leq 3.16 \times 10^{-3} \Msun \rm pc^{-3}$, which corresponds 
to the critical density for star formation.

In Figure~\ref{fig:xray_cosmo}, we show the X-ray luminosity of the 
hot gas against (a) the stellar velocity dispersion $\sigma$, (b) total 
mass within $5 \re$, and (c) virial mass $\Mvir$ of the simulated 
galaxies at $z=0$ for the simulated central galaxies with three 
feedback models. Observed relations for the normal early-type
galaxies are from (a) \citet{2011ApJ...729...12B}, 
(b) \citet{2013ApJ...776..116K}, and (c) \citet{2006ApJ...652L..17M} 
respectively. Observationally all galaxies with a shallow potential 
well with $\sigma < 200~\kms$ seem to have only a small amount of 
hot gas with $L_X < 10^{40}~\ergs$ \citep{2011ApJ...729...12B}, 
however, ThAGN models show much higher X-ray luminosity for up 
to $\sim$ 2-3 orders of magnitude higher. On the other hand, MrAGN 
models produce X-ray luminosity within the observed range, with the 
two orders of magnitude spread in $L_X$. Note that we obtain higher 
X-ray luminosity $L_X$ for ThAGN models compared to the work by 
\citet{2006ApJ...643..692C}, who studied the X-ray luminosity and the 
scaling relation of merger remnants. This discrepancy seems to be 
due to the differences in X-ray bandwidths, the inclusion of metal-line 
emission and the SPH formulation. We use the X-ray luminosity in 
{\it Chandra} bands (0.3-8 keV) including bremsstrahlung radiation 
and metal-line emission from all relevant species, while
\citet{2006ApJ...643..692C} used bolometric X-ray luminosity and 
partially included metal line emission (for 0.1 to 2 KeV). 
\citet{2006ApJ...643..692C} used the standard formation of SPH 
which tends to suppress mixing and cooling of the hot gas and result 
in lower X-ray luminosity.

\subsection{Baryonic mass budget}
\begin{figure}
\epsfig{file=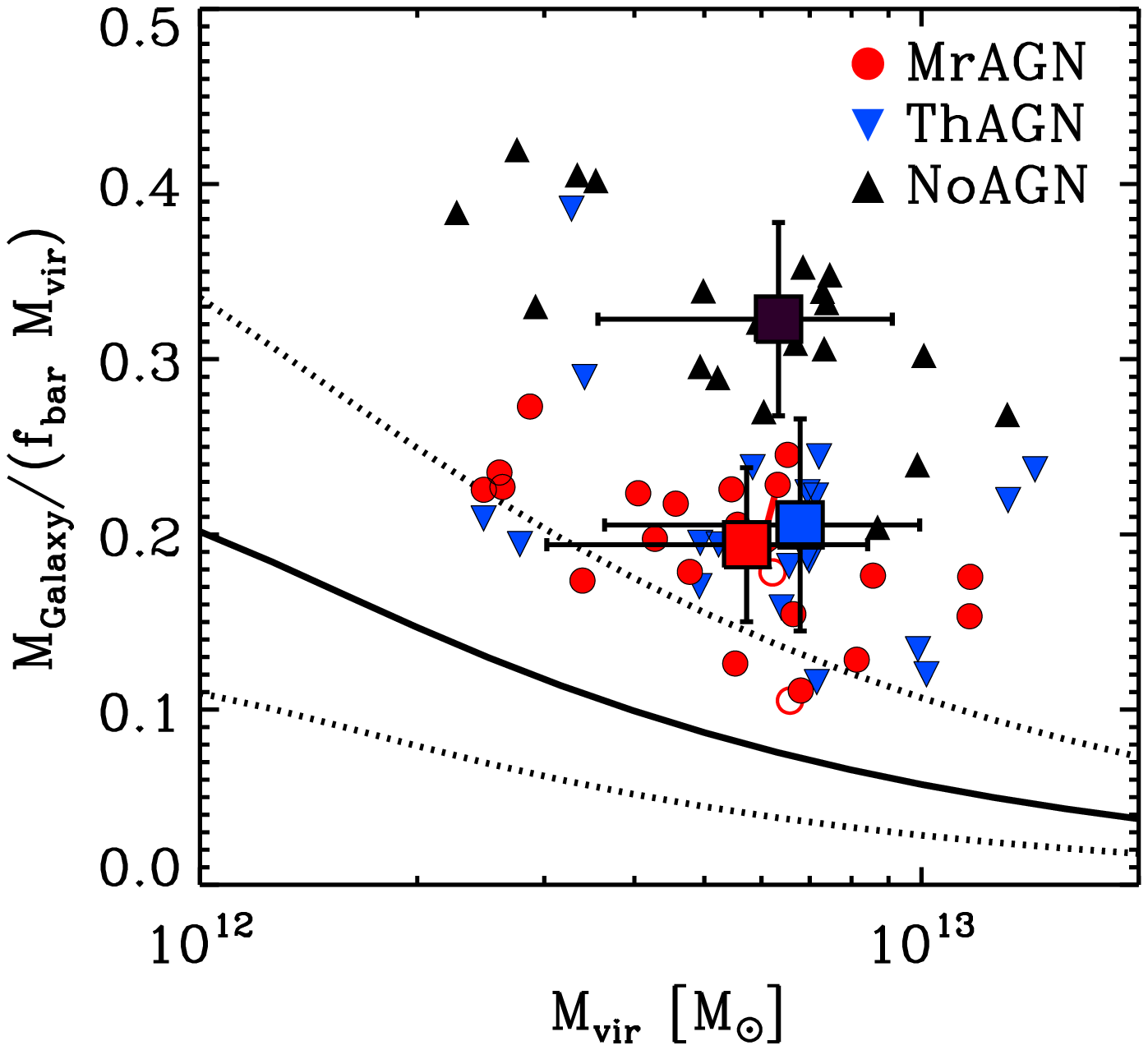,width=\columnwidth}
\caption{Fraction of baryons that is converted into stars at $z=0$ for 
the model without AGN feedback (NoAGN, black triagles), thermal 
feedback (ThAGN, blue upside down triangles), and momentum and 
radiative feedback (MrAGN, red circles). The momentum feedback 
runs without radiative feedback are shown in open red circles for two 
halos. Respective mean values  (and their scatter) are indicated by 
the squares. The AGN feedback models have lower conversion 
efficiencies than the NoAGN model but are still a factor of 2 higher 
than abundance matching estimates from 
\citet{2013MNRAS.428.3121M} (solid black line).  
\label{fig:baryonic}}
\end{figure}

\begin{figure*}
\begin{center}
\epsfig{file=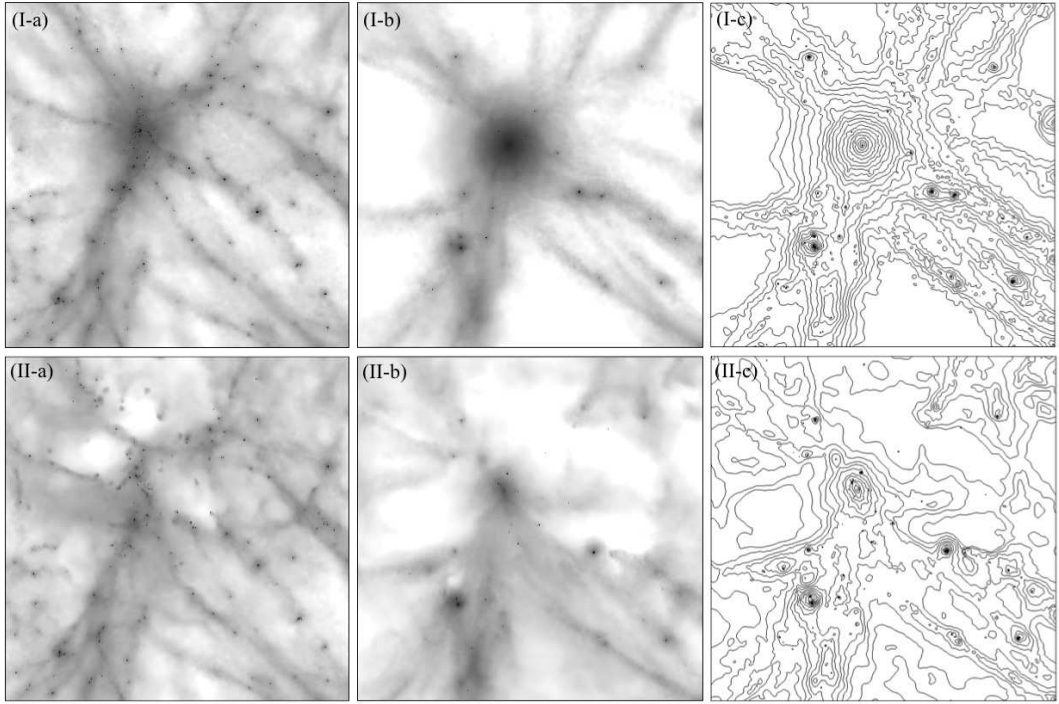,width=0.8\textwidth}
\caption{Projected gas density maps for a typical central galaxy with 
a halo mass of $5 \times 10^{12} \Msunh$ for the traditional thermal 
AGN feedback in the top panels, and for the momentum AGN 
feedback in the bottom panels. The projected gas density maps at 
$z=1.5$ are shown in the left-most panels (I-a, II-a), and maps at 
$z=0$ are shown in the middle panels (I-b, II-b). Coutours of 
projected gas density at $z=0$ are shown in right-most panels 
(I-c, II-c). The images are 4 Mpc on a side and brighter colour 
indicates a lower density. While the ThAGN model does not generate 
any high velocity outflows and bubbles as shown in upper panels, in 
the MrAGN model the high velocity winds with $\vw \sim 10,000$ 
$\kms$ effectively shock the ambient gas and generate cavities 
around the central massive galaxies in panel (II-a). Bubbles further 
expand and effectively lower the gas density around the central 
elliptical galaxy in panels (II-b) and (II-c).
\label{fig:map}}
\end{center}
\end{figure*}

\begin{figure*}
\begin{center}
\epsfig{file=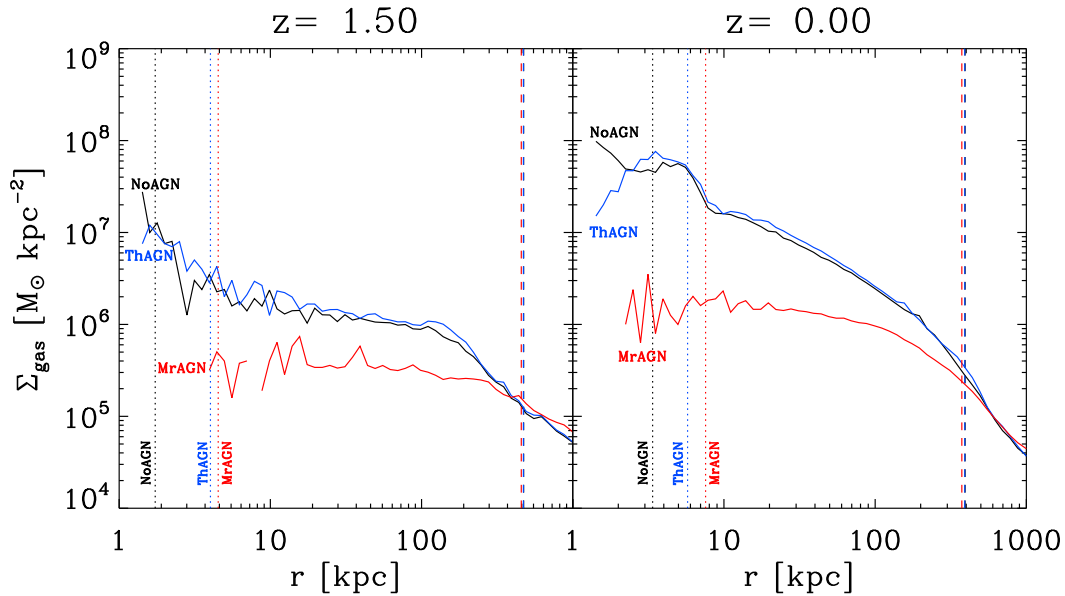,width=0.85\textwidth}
\caption{Projected gas density profiles of the central galaxies in a 
zoom-in simulation of a $5 \times 10^{12} \Msunh$ dark matter halo 
at $z=1.5,0$ (left and right panel) for the model without AGN feedback 
(NoAGN, black), thermal feedback (ThAGN, blue), and momentum 
feedback (MrAGN, red). The virial radius $\rvir$ and effective radius 
$\re$ of each model are shown in dashed lines and dotted lines 
respectively. The gas density in the momentum feedback model 
(MrAGN) is already lower than the others at $z=1.5$ in all radii within 
$\rvir$, and the central density  becomes $\sim 20$ times lower at 
$z=0$.  The thermal feedback model (ThAGN) and the model without 
AGN feedback (NoAGN) show very similar profiles at z=1.5, but at 
$z=0$ ThAGN shows lower density than NoAGN for $r < 2$ kpc.
\label{fig:gasradial}}
\end{center}
\end{figure*}
 
Fig. \ref{fig:baryonic} shows the conversion efficiencies of the 
simulated galaxies at redshift zero $f_* = m_*/(f_b*m_{vir,dark})$ 
where $m_*$ is the stellar mass within 10 \% of the virial radius 
$\r10$, $f_b=\Omega_b/\Omega_{dm} =0.17$ is the cosmic baryon 
fraction and $m_{vir,dark}$ is the dark matter mass within the virial 
radius of the galaxy. The amount of total baryonic matter available in 
each halo is $f_b * m_{vir,dark}$ and then we calculate $f_*$, the 
fraction thereof that is converted into stars in the simulated central 
galaxy. For simulated galaxies with NoAGN model, the fraction 
declines in a roughly linear fashion with the logarithm of the halo 
mass from $f_* \approx 0.6$ for the smallest halos 
($\approx 10^{12} M_{\odot}$) to $f_* \approx 0.2$ for high mass 
halos ($\gtrsim 10^{13} M_{\odot}$), over-predicting by a factor of 
$\sim 2-3$ than the estimates from recently published work of the 
conversion efficiencies based on the abundance matching technique 
\citep{2013MNRAS.428.3121M}. Note that the conversion 
efficiencies of NoAGN models we obtain in this study are slightly 
higher than the value from \cite{2012ApJ...744...63O}. Using the 
modified SPH version which allows for more efficient mixing we find 
that more gas cools resulting in higher conversion efficiencies. In 
case of AGN feedback models, both ThAGN and MrAGN models 
show lower conversion efficiencies compared to NoAGN models, but 
they still over-predict by a factor of $\sim 2$ than the estimates and 
are expected to further increase if metal-line cooling was included. 
This seems to be due to the absence of the ejective supernova wind 
feedback, which is especially important for low-mass systems. Our 
simulations do not generate significant winds in the low-mass 
systems which do not have massive black holes, and thus 
overestimate the condensed baryon fraction of low-mass galaxies 
that can accrete onto high mass galaxies at late stages of evolution. 
This may produce a factor of 2 discrepancy for high-mass galaxies 
because of over-predicted accreted stars from the low-mass galaxies. 
The discrepancy is expected to be effectively reduced by including 
wind stellar feedback, as the work by \citet{1991ApJ...376..380C}
has shown that Type I SN in normal ellipticals is capable of ejecting 
significant fraction of the secondary gas from the systems. In recent 
study by \cite{2013MNRAS.436.2929H}, galactic winds driven by SN 
are found to be effective in suppressing central star formation in less
massive systems.

\subsection{Gas mass fraction and Star formation rate}
One of the most important findings in the MrAGN feedback models 
is the effective AGN-driven quenching of star formation in central 
galaxies in consistent with the observation 
\citep[e.g.][]{2012Natur.485..213P}. The high velocity outflowing 
winds shock the ambient gas and generate cavities around the 
central massive galaxies, significantly lowering the gas density and 
suppressing star formation in MrAGN model. 

To illustrate this effect, in Figure~\ref{fig:map} we show the projected 
gas density maps of the central galaxies at $z=1.5,0$ in a zoom-in 
simulation of a $5 \times 10^{12} \Msunh$ dark matter halo with the 
ThAGN feedback, and with the MrAGN feedback. While the ThAGN 
model does not generate any high velocity outflows and bubbles, in 
the MrAGN model the high velocity winds with 
$\vw \sim 10,000$~$\kms$ effectively shock the ambient gas and 
generate cavities around the central massive galaxies. Bubbles 
further expand and effectively lower the gas density around the 
central elliptical galaxy.

The gas density lowered by the AGN wind is quantitatively shown in 
Figure~\ref{fig:gasradial}. It shows the projected gas density profiles
of the central galaxies at $z=1.5,0$ in a zoom-in simulation of a 
$5 \times 10^{12} \Msunh$  dark matter halo, identical galaxies 
shown in Figure~\ref{fig:map}. The gas density in the momentum 
feedback model (MrAGN) is already lower
than the others at $z=1.5$ in all radii within $\rvir$, and the central 
density becomes $\sim 20$ times lower at $z=0$.  The thermal 
feedback model (ThAGN) and the model without AGN feedback 
(NoAGN) show very similar profiles at z=1.5. At $z=0$ ThAGN shows 
lower density than NoAGN but only for $r < 2$ kpc. The ThAGN 
models have localized effects within  $r<2$ kpc.

\begin{figure}
\epsfig{file=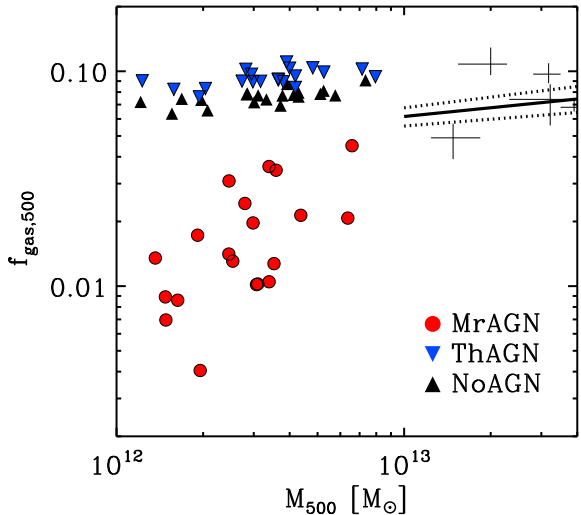,width=\columnwidth}
\caption{The enclosed gas mass fraction within $r_{500}$, 
$f_{\rm gas,500} - M_{500}$ relation for the different AGN feedback 
models: NoAGN in black triangles, ThAGN in blue upside down 
triangles and MrAGN in red circles. The black crosses and black 
solid lines are the observed data and fitted 
$f_{\rm gas,500} - M_{500}$ relation of the nearby galaxy groups of 
$M_{500}=10^{13-14} \Msun$ from \citet{2009ApJ...693.1142S}.
\label{fig:fgas}}
\end{figure}

\begin{figure}
\epsfig{file=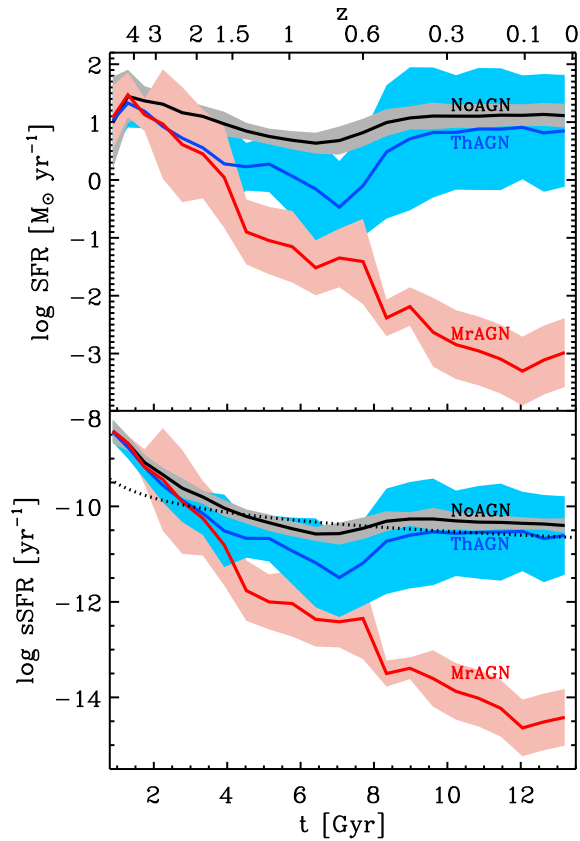,width=\columnwidth}
\caption{Top:  Averaged star formation rate over time for the central 
galaxies for the model without AGN feedback (NoAGN, black), 
thermal feedback (ThAGN, blue), and momentum and radiative 
feedback (MrAGN, red) measured within the 10 \% of the virial radii. 
The median star formation rate for each model are shown in solid 
lines and the coloured areas show the 1$\sigma$ scatter of the mean 
star formation rates. Bottom: same as in the top panel but for the 
median specific star formation rates. The dotted line indicates the 
specific star formation rates equal to~$ 0.3/\tH$, commonly adopted 
criteria separating quiescent and star forming galaxies 
\citep[e.g.][]{2008ApJ...688..770F}. Sources above this line are in a 
star forming mode: NoAGN feedback model (black) stays above this 
line through out. Star formation in ThAGN model (blue) is reduced 
below $ 0.3/\tH$ at $z \sim 1.5$ but increases again as gas cools and 
falls back again. In the MrAGN feedback model (red), the AGN 
feedback drives a large-scale wind and effectively quenches star 
formation.
\label{fig:sfr}}
\end{figure}
 
In Figure~\ref{fig:fgas} we compare the gas mass fraction of all 
simulated halos to observation. We measure the fraction of gas mass 
to the total mass within $r_{500}$, the radius where the spherical 
over-density drops below 500 times the critical density of the 
universe at $z=0$. The observed $f_{\rm gas,500}-M_{500}$ relation 
derived from 43 nearby galaxy groups with $ M_{500} = 10^{13-14} 
\Msun$ from \citet{2009ApJ...693.1142S} is shown in black solid line 
with 1$\sigma$ scatter in dotted lines. We also overplot the observed 
gas mass fraction of galaxy groups from \citet{2009ApJ...693.1142S}
in black crosses. In MrAGN model, the strong wind results in much 
lower gaseous mass fraction compared to other feedback models 
especially for the low mass galaxies. The differences between 
MrAGN and other feedback models becomes smaller as MrAGN 
model show higher gaseous mass fraction in higher mass galaxies. 
This is consistent with the observed relation which shows a strong 
trend in gas mass fraction with total mass, such that galaxy groups 
have significantly lower fractions compared to massive clusters and 
the universal baryon fraction $f_b = 0.17$. 
\citet{2011MNRAS.412.1965M,2014MNRAS.441.1270L} also 
previously showed that  a model with AGN feedback can reproduce 
X-ray luminosity and gas mass fractions of groups and clusters of 
galaxies.
 
In Figure~\ref{fig:sfr} we show the median star formation rate over 
time with the various feedback models, considering stars within the 
10\% of the virial radius ($\r10$) of the galaxy centre. In MrAGN 
model (red), the AGN feedback drives a large-scale wind that 
removes nearly all of the residual gas from the galaxy, effectively 
and rapidly quenching star formation. Compared to the commonly 
adopted criteria for star forming galaxies, sSFR $> 0.3/\tH$, where 
$\tH$ is the age of the universe at each redshift \citep[e.g.][marked 
with dotted line in the bottom panel]{2008ApJ...688..770F}, MrAGN 
model galaxies can be considered quiescent since $z \sim 2$. When 
AGN feedback is not included, central galaxies keep star-forming
and they are classified as star-forming galaxies throughout the 
simulations. In ThAGN model, in-situ star formation is quenched 
around $z=1$, but the star formation rate increases again as cooled 
gas fall back to the central region. The late time star formation rate is 
seen to be more than two orders of magnitude lower in the MrAGN 
than in the ThAGN model. The MrAGN model will be further tested 
by comparing the star formation activities of the larger sample of 
simulated galaxies to observations in a forthcoming paper.

\subsection{Galaxy sizes and velocity dispersions}
\begin{figure}
\epsfig{file=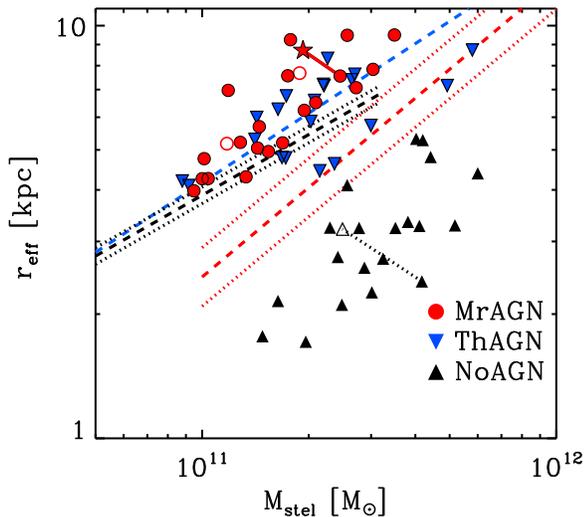,width=\columnwidth}
\caption{Projected stellar half-mass radii of the simulated galaxies 
versus stellar masses for redshifts z=0 for the model without AGN 
feedback (NoAGN, black triangles), thermal feedback (ThAGN, blue 
upside down triangles), and momentum and radiative feedback 
(MrAGN, red circles). The blue/black/red dashed lines indicate the 
observed size--mass relation for early-type galaxies respectively from 
Shen et al. (2003), Hyde \& Bernardi (2009), and Nipoti et al. (2009) 
with the $1 \sigma$ scatter indicated by the dotted lines. The 
momentum feedback runs without radiative feedback are shown in 
open red circles for two halos. Open black triangle shows the 
half-mass radius of the simulated galaxy with a density-dependent 
SPH formulation, consistent with Oser et al. (2012).
\label{fig:size}}
\end{figure}

\begin{figure}
\epsfig{file=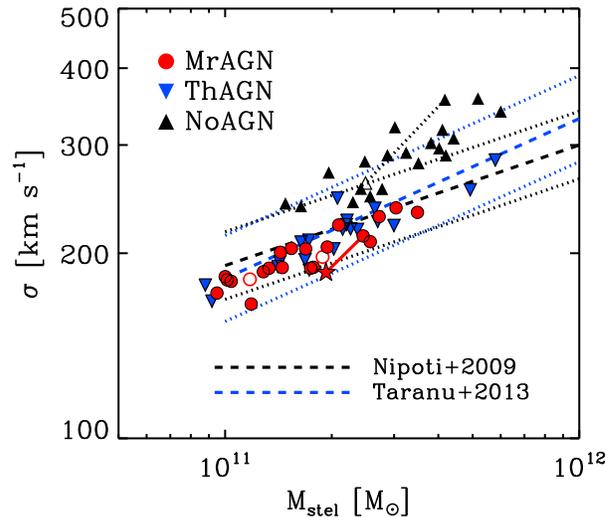,width=\columnwidth}
\caption{Central (within 0.5 $r_{\rm eff}$) projected velocity dispersion
as a function of stellar masses at z=0  for the model without AGN 
feedback (NoAGN, black triangles), thermal feedback (ThAGN, blue 
upside down triangles), and momentum and radiative feedback 
(MrAGN, red circles). The momentum feedback runs without radiative 
feedback are shown in open red circles for two halos. The black/blue 
dashed lines indicate the observed relation for early-type galaxies 
from Nipoti et al. (2009) and Taranu et al. (2013) respectively. We 
indicate the observed $1 \sigma$   scatter with the dotted lines. Open 
black triangle shows the velocity dispersion of the simulated galaxy 
with a density-dependent SPH formulation, consistent with Oser et al. 
(2012).
\label{fig:sigma}}
\end{figure}

In Figure~\ref{fig:size}, we show the projected half-mass radii of 
simulated galaxies of MrAGN model as a function of their stellar 
masses at $z=0$.  We determine the effective radius of the galaxy 
$\re$ by determining the mean values of the half-mass radii of all 
stars within 10$\%$ of the virial radius $\r10 = 0.1 \times \rvir$ 
projected along the three principal axes of the main stellar body. The 
blue/black/red dashed lines indicate the observed size--mass relation 
for early-type galaxies respectively from 
\cite{2003MNRAS.343..978S}, \cite{2009MNRAS.394.1978H}, and 
\cite{2009ApJ...706L..86N}, with the $1 \sigma$ scatter indicated by 
the dotted lines. We show the half-mass radius and the stellar mass 
of the simulated galaxies with a normal density-dependent SPH
formulation, consistent with \citet{2012ApJ...744...63O}, in open
black triangle. The corresponding NoAGN models simulated with the 
new pressure-entropy formulation are connected with black dotted 
lines.  The SPHGal with improved mixing typically reduces the sizes 
by on average $\sim$30 percent and also increases the stellar 
masses compared to normal SPH runs. When AGN feedback is 
included there is a factor of 2--3 increase in size compared to the 
NoAGN models.

Figure~\ref{fig:sigma} shows the relation between the velocity
dispersion $\sigmastar$ of simulated galaxies and their stellar 
masses \citep[Faber-Jackson relation,][]{1976ApJ...204..668F} for 
MrAGN model at $z=0$. We obtain the line-of-sight velocity 
dispersions $\sigmastar$ within $0.5 \times \re$ along the three 
principal axes and compare to observations. The observed relations 
are from \cite{2009ApJ...706L..86N} who used the SLACS sample of 
local early-type galaxies at $z=0$ and \cite{2013ApJ...778...61T} who 
analyzed \citet{2010ApJS..186..427N} SDSS catalog. Compared to 
the observation,  ThAGN and MrAGN models predict reasonable 
velocity dispersions  in agreement with \cite{2012ApJ...744...63O}.

\section{Summary}\label{summary_cosmo}
In this paper, we have run three sets of cosmological hydrodynamic
simulations of 20 halos with masses between $2.3 \times 10^{12}
\Msun$ $ \lesssim \Mvir \lesssim 1.4 \times 10^{13} \Msun$ using a 
modified GADGET-3 SPH code with different AGN feedback models: 
(1) No black hole and AGN feedback (NoAGN), (2) the standard 
thermal AGN feedback (ThAGN) \citep[e.g.][]{2005Natur.433..604D,
2005MNRAS.361..776S,2007MNRAS.380..877S,
2008ApJ...676...33D} and (3) mechanical/radiation AGN feedback 
(MrAGN) \citep{2012ApJ...754..125C,2014MNRAS.442..440C}.
We study the effects of AGN feedback on the black-holes and the 
properties of their host galaxies, i.e. black-hole scaling relation 
($\Mbh-\sigmastar$ and $\Mbh/\Mstel$), X-ray luminosity, stellar and 
gaseous baryonic conversion efficiencies, star formation, and size of 
the elliptical galaxies. 

We show that massive, non-relativistic outflows and X-ray heating in
mechanical and radiation AGN feedback model  indeed provide a 
viable mechanism to regulate the black hole growth in central early 
type galaxies. The observed $\Mbh - \sigma$ relationship between 
the black hole mass and the galaxy velocity dispersion is 
successfully recovered with the MrAGN model. This was also 
obtained in the previous ThAGN feedback treatments 
\citep[e.g.][]{2005Natur.433..604D,2005MNRAS.361..776S}, and we 
confirm their results on the observed physical relation between black 
hole and galactic properties, with our modified SPH code and 
cosmological simulations. Most importantly, however, the MrAGN 
model shows much lower X-ray luminosity compared to the 
commonly adopted ThAGN model where all the feedback energy is 
distributed locally as thermal heating. While the thermal feedback 
model produces $\sim$2-3 orders of magnitude higher thermal X-ray 
luminosity than expected for given stellar mass of the galaxy, the 
MrAGN model can successfully reproduce both the observed 
$L_{X}-\sigmastar$ and $\Mbh-\sigmastar$ relations.

We show that mechanical and radiation feedback can also effectively 
suppress in-situ star formation in high mass galaxies as the gas is 
effectively expelled by AGN-driven outflows especially at late times. 
The MrAGN model produces massive red and dead galaxies,
reducing final stellar mass by a factor of two compared to the 
previous works without AGN feedback \citep{2010ApJ...725.2312O,
2012ApJ...744...63O} and reducing the final star formation rates by 
over $10^3$ compared with the no AGN feedback or thermal AGN 
feedback models. Our simulated galaxies with AGN feedback show 
lower baryonic conversion efficiencies than the ones without AGN 
feedback, but our samples are still overly efficient in transforming
gas into stars. The baryonic conversion efficiencies at $z=0$ in our 
simulated galaxies with the MrAGN feedback are still overestimated 
by roughly a factor of $\sim 2$ compared to recent predictions 
\citep{2013MNRAS.428.3121M} and would be even enhanced if 
metal-line cooling was included. As AGN feedback mainly affects 
massive galaxies with massive black holes and quenches star 
formation in the inner region, `in-situ' star formation is significantly 
suppressed in our models resulting lower fraction of in-situ stars 
compared to the previous models without AGN feedback. One or 
more missing further mechanisms is needed to reduced the star
formation in low mass satellite galaxies that are eventually accreted., 
e.g. supernova wind feedback \citep{2005ApJ...618..569M,
2006MNRAS.373.1265O,2012MNRAS.426..237H,
2013MNRAS.436.2929H,2013MNRAS.428.2966P}, cosmic ray 
driven winds \citep{2013ApJ...777L..16B,2013ApJ...777L..38H,
2014MNRAS.437.3312S}, and/or star formation driven wind 
\citep{2014arXiv1404.2613A,2014MNRAS.445..581H}. The late time 
quenching can be more challenging for both ThAGN and MrAGN 
models as the material removed from lower mass progenitors can be 
accreted to more massive ones at later times 
\citep{2010MNRAS.406.2325O}.

\section*{Acknowledgments}
We benefited from useful conversations with
Renyue Cen, Taysun Kimm, Rachel Somerville, James M. Stone, and Michael A. Strauss. 
We also thank the anonymous referee for many constructive 
suggestions, which helped to improve the presentation of the results.
E.C. and J.P.O. acknowledge the support of NSF grant 
AST-0707505. E.C. and T.N. acknowledge the support 
from the DFG cluster of excellence ``Origin and Structure of the 
Universe''.  E.C. was supported by the Samsung Scholarship 
foundation and made extensive use of the computing facilities of 
the Princeton Institute of Computational Science and Engineering.

\bibliography{references}
\end{document}